\documentclass[superscriptaddress]{revtex4}
\usepackage{amsmath}
\usepackage{amsfonts}
\usepackage{amssymb}
\usepackage{graphicx}

\begin{document}

\title{Spatial entanglement of bosons in thermal equilibrium}
\author{Dagomir Kaszlikowski}\affiliation{Departament of Physics, National University of Singapore, 117542 Singapore, Singapore}
\author{Andreas Keil}\affiliation{Departament of Physics, National University of Singapore, 117542 Singapore, Singapore}
\author{Marcin Wie\'sniak}\affiliation{Institute of Theoretical Physics and
Astrophysics, University of Gda\'nsk, PL-80-952 Gda\'nsk,
Poland}\affiliation{Departament of Physics, National University of
Singapore, 117542 Singapore, Singapore}
\author{Frederick H. Willeboordse}\affiliation{Departament of Physics, National
University of Singapore, 117542 Singapore, Singapore}
\begin{abstract}
We investigate spatial entanglement in a system of
spinless non-interacting bosons in a one dimensional harmonic trap described by the grand canonical ensemble. We find the lower bound for the amount of spatial entanglement contained in the average state of two bosons in the trap and show that it exists for arbitrary high temperatures.
\end{abstract}
\maketitle
\section{Introduction}

Since its discovery, the phenomenon of entanglement lacks full
understanding. For instance, when one deals with mixed states, we do
not have necessary and sufficient criteria to determine if a given
state is entangled or separable. One has, of course, numerous
criteria for entanglement, for example the partial transposition
test \cite{pt}, or the reduction criterion \cite{red}, but they
provide only necessary conditions. Mixed states, on the other hand,
are commonly observed, as quantum systems interact with their
environment. In particular, a natural state of the matter is a thermal
mixture. Therefore, the question whether a (thermal) density matrix
is entangled or not, and how entanglement is related to
thermodynamical principles and quantities can arise in many physical
situations. The notion of entanglement becomes even less clear when the effects
of Bose-Einstein (BE) or Fermi-Dirac (FD) statistics cannot be neglected. This
is the case when the wave functions of particles forming a physical
system significantly overlap (as in thermal mixture). This problem has recently been
investigated in Ref. \cite{p1,p2,p3,p4,p5,p6,p7}.

There are many interesting physical situations where BE and FD
statistics play an important role. Standard examples are fermionic
and bosonic gases, quantum phase transitions \cite{qpt},
super-conductivity \cite{super-con}, super-fluidity
\cite{super-fluid} and
Bose-Einstein Condensation (BEC) \cite{einstein}, a phenomenon in
which a large group of indistinguishable bosons is described by the
same single-boson ground-state wave function. It is interesting to
see whether entanglement is related to these phenomena and to what
extent. Concerning BEC, this problem was addressed for zero
temperature in the Ref. \cite{simon} and for non-zero temperature in the Ref. \cite{anders}.

In this paper we discuss the so-called spatial entanglement
\cite{anders} in the simplest model of non-interacting spinless
bosons trapped in a one-dimensional harmonic oscillator potential
described by the grand canonical ensemble. In this example BE
statistics cannot be neglected and the system exhibits phase
transition in the form of Bose-Einstein condensation
\cite{ketterle}.

\section{Spatial entanglement}
We start our discussion by defining what we understand by spatial
entanglement in a system of indistinguishable non-interacting bosons. We limit our discussion to a one dimensional case but
the extension to two and three dimensions is straightforward.

Let us first consider the simples possible case of two bosons moving in some trapping potential. The eigenkets of the Hamiltonian corresponding to eigenvalues $E_k$ are given by $|\phi_k\rangle$ ($k=0,1,2,\dots$) and the symmetric basis for two bosons reads $|\psi\rangle_{k,l} = \frac{1-\delta_{kl}}{\sqrt{2}}(|\phi_k\rangle|\phi_l\rangle+|\phi_l\rangle|\phi_k\rangle)+\delta_{kl}|\phi_{k}\rangle|\phi_{k}\rangle$. Are these basis states entangled? This question does not make sense unless we precisely define what the degrees of freedom are in relation to which we discuss entanglement. Additionally, bosons are indistinguishable, which further complicates the answer.

To see the problem more clearly, let us consider a state describing two bosons that are spatially separated, i.e., their respective wave functions do not overlap. A simple example is given by two indistinguishable bosons each confined in its own infinite square well potential, say $A$ and $B$. The eigenstates of the wells $A$ and $B$ are $|\phi^A_k\rangle$ and $|\phi^B_l\rangle$ respectively. It is possible to prepare the particles in the wells in the state
$\frac{1}{\sqrt 2}(|\phi^A_k\rangle |\phi^B_l\rangle+|\phi^A_l\rangle |\phi^B_k\rangle)$ with $k\neq l$. This state is, without doubt, entangled since the the partially transposed density matrix with respect to the subsystem $B$ has negative eigenvalues. Here there is no ambiguity which degrees of freedom (subsystems) are entanglement or in other words what the subsystems $A$ and $B$ are. We can unambiguously identify such subsystems because the two bosons are spatially separated and become physically distinguishable in the sense that local measurements performed on them will not reveal the bosonic properties of the particles.

Now, let us put these two bosons into the same trapping potential such that their energies are $E_k$ and $E_l$ with $k\neq l$. The state describing this situation is $\frac{1}{\sqrt 2}(|\phi_k\rangle |\phi_l\rangle+|\phi_l\rangle |\phi_k\rangle)$ and it looks deceptively similar to the previous one. However, now there is nothing that makes these two bosons different or, as opposed to the previous example, there are no well-defined subsystems $A$ and $B$. We now investigate what happens to entanglement in this case.

We could naively define subsystems $A$ and $B$ by analogy with the
previous example, i.e., the first ket would correspond to $A$ and
the second one to $B$. Then we could perform partial transposition
with respect to $A$ and check for the negativity of the density
matrix obtained this way. We would get exactly the same result as in
the previous example. However, this operation is physically
meaningless because the result we obtain is a negative density
matrix that is not symmetric with respect to a relabeling of the
particle indices. This is a serious problem because partial
transposition corresponds to the reversal of time on one of the
subsystems, which cannot change the fact that we deal with bosons.
This in turn means that even the unphysical negative operator after
partial transposition must be still symmetric. We did not have this
problem with the previous example where the bosonic character of the
particles was irrelevant due to the physically meaningful labels
distinguishing them. The labels enumerating bosons in this example
do not have any physical meaning (there is no boson one and two) and
one has to find other means to distinguish them. We have mentioned
this example in relation to the Ref. \cite{wrong} where the
criterion of entanglement for two bosons in a harmonic trap was
based on the partial transposition with respect to the indices
enumerating bosons. In view of the above arguments the results in
the Ref. \cite{wrong} are not correct.

In the following part of this section we define physically meaningful subsystems (first for two then for many bosons) $A$ and $B$ and discuss entanglement between them. In order to distinguish this type of entanglement from other types of entanglement we henceforth refer to it as \textit{spatial  entanglement}. Our definition is not unique and one could conceive other ways of defining entanglement \cite{Zanardi}. The reason why we choose to investigate the spatial entanglement defined in this way is that it may be related to the phenomenon of off-diagonal long range correlations, which in turn is related to phase transitions \cite{Yang}.

First we define two non-overlaping adjacent intervals dividing the whole space into two parts, say $A$ and $B$. With respect to this division the eigenket $|\phi_k\rangle$ in the position representation can be now written as $\langle x|\phi_k\rangle =\phi_k(x) = \sqrt{p_A(k)}\phi^A_k(x) + \sqrt{p_B(k)}\phi^B_k(x)$, where $\sqrt{p_{A(B)}(k)}$ is the probability of finding the particle in the region $A$ and $\phi^{A(B)}_k(x)$ is the wave function defined as $\frac{1}{\sqrt{p_{A(B)}(k)}}\phi_k(x)\chi_{A(B)}(x)$. Here $\chi_{A(B)}(x)$ is a characteristic function of the region $A(B)$, i.e., it takes the value $1$ if $x$ belongs to $A(B)$ and $0$ otherwise. Similarly, the symmetrized state $\psi_{kl}(x_1,x_2)$ of two bosons (in position representation) can be written as
\begin{equation}
\psi_{kl}(x_1,x_2) = \sqrt{p_{AA}(kl)}\psi^{AA}_{kl}(x_1,x_2)+\sqrt{p_{BB}(kl)}\psi^{BB}_{kl}(x_1,x_2)+\sqrt{p_{AB}(kl)}\psi^{AB}_{kl}(x_1,x_2),
\label{state}
\end{equation}
where
\begin{eqnarray}
&&\psi^{AA}_{kl}(x_1,x_2)=\frac{1}{\sqrt{p_{AA}(kl)}}\psi_{kl}(x_1,x_2)\chi_A(x_1)\chi_A(x_2)\nonumber\\
&&\psi^{BB}_{kl}(x_1,x_2)=\frac{1}{\sqrt{p_{BB}(kl)}}\psi_{kl}(x_1,x_2)\chi_B(x_1)\chi_B(x_2)\nonumber\\
&&\psi^{AB}_{kl}(x_1,x_2)=\frac{1}{\sqrt{p_{AB}(kl)}}\psi_{kl}(x_1,x_2)(\chi_A(x_1)\chi_B(x_2)+\chi_A(x_2)\chi_B(x_1)).
\end{eqnarray}
Obviously, $p_{AA}(kl)$ denotes the probability of finding two bosons in $A$, $p_{AB}(kl)$ denotes the probability of finding one of the bosons in $A$ and the other one in $B$ etc. These probabilities can easily be found by computing the appropriate integrals. All the above wave functions are symmetric and they describe sub-ensembles corresponding to three physically distinguishable situations; two bosons in the region $A$, two bosons in the region $B$ and bosons in separate regions.

Let us now discuss the spatial entanglement between these two regions. We can compute the amount of entanglement contained in the state $|\psi_{kl}\rangle$ by computing the entropy of the subsystem $A$, which results from tracing out the subsystem $B$ (this is a good measure because we deal with pure bipartite states). This is a similar approach to that in Ref. \cite{simon}.

Before computing the von Neumann entropy of the subsystem $A$ let us write the state (\ref{state}) without using the position representation. The notation we adopt here is similar to that of the occupation number representation in the so-called second quantization. We consider the two cases $k\neq l$ and $k=l$ separately. For $k\neq l$ we have
\begin{eqnarray}
&&|\psi_{kl}\rangle = \sqrt{p_{AA}(kl)} |1_k,1_l\rangle_A |0\rangle_B +\sqrt{p_{BB}(kl)} |0\rangle_A |1_k,1_l\rangle_B+\nonumber\\
&&(\sqrt{p_A(k)p_B(l)}|1_k\rangle_A|1_l\rangle_B+\sqrt{p_A(l)p_B(k)}|1_l\rangle_A|1_k\rangle_B),
\end{eqnarray}
where $|0\rangle_{A(B)}$ denotes the vacuum in the region $A(B)$.
This is not a physical vacuum like in field theory, it is simply a
ket conveying the information that there are no particles in $A(B)$.
The state $|1_k,1_l\rangle_{A(B)}$ means that there are two bosons
in the region $A(B)$ one of them in the state
$|\phi^{A(B)}_k\rangle$ and the second one in the state
$|\phi^{A(B)}_l\rangle$. The states $|1_k\rangle_A|1_l\rangle_B$ and
$|1_l\rangle_A|1_k\rangle_B$ describe the situation when one of the
bosons is in the region $A$ and the second one in the region $B$ in
the appropriate states $|\phi^{A(B)}_{k(l)}\rangle$. More precisely
\begin{eqnarray}
&&|1_k,1_l\rangle_{A(B)} = \frac{1}{\sqrt{p_{AA(BB)}(kl)}}(|\phi^{A(B)}_k\rangle |\phi^{A(B)}_l\rangle+|\phi^{A(B)}_l\rangle |\phi^{A(B)}_k\rangle)\nonumber\\
&&|1_l\rangle_{A}|1_k\rangle_B = \frac{1}{\sqrt{2}}(|\phi^{A}_l\rangle|\phi^B_k\rangle+|\phi^B_k\rangle |\phi^A_l\rangle).
\end{eqnarray}
It is important to stress that the normalized states $|1_k,1_l\rangle_{A(B)}$ and $|1_k\rangle_{A(B)}$ {\it do not} form an orthogonal basis in the two- and single-particle subspace respectively in the Hilbert space describing the sector $A(B)$. We also note that these states are symmetric with respect to exchange of fictitious indices enumerating bosons.

It is interesting to write explicitly the probabilities appearing in the above formulas
\begin{eqnarray}
&&p_{AA}(kl) = p_A(k)p_A(l)(1+|{}_A\langle 1_k|1_l\rangle_A|^2)\nonumber\\
&&p_{BB}(kl) = p_B(k)p_B(l)(1+|{}_B\langle 1_k|1_l\rangle_B|^2)\nonumber\\
&&p_{AB}(kl) = 1-p_{AA}(kl)-p_{BB}(kl).
\end{eqnarray}
We observe a bunching effect since the probability of finding two bosons either on the left or on the right is enhanced in comparison to distinguishable particles for which the respective probabilities would be $p_{AA}(kl) = p_A(k)p_A(l)$ and
$p_{BB}(kl) = p_B(k)p_B(l)$. This enhancement means that the probability of anti-bunching, i.e., finding each boson in a separate region is reduced when compared to distinguishable particles. The effect disappears if the overlaps $|{}_{A(B)}\langle 1_k|1_l\rangle_{A(B)}|$ are zero. If $k=l$ the formulas simplify considerably and we have
\begin{equation}
|\psi_{kk}\rangle = p_{A}(k)|2_k\rangle_A|0\rangle_B+p_{B}(k)|0\rangle_A|2_k\rangle_B+\sqrt{2 p_A(k)p_B(k)}|1_k\rangle_A |1_k\rangle_B,
\label{exotickk}
\end{equation}
where
\begin{eqnarray}
&&|2_k\rangle_{A(B)} = |\phi^{A(B)}_k\rangle|\phi^{A(B)}_k\rangle\nonumber\\
&&|1_k\rangle_A |1_k\rangle_B = \frac{1}{\sqrt{2}}(|\phi^A_k\rangle|\phi^B_k\rangle+|\phi^B_k\rangle|\phi^A_k\rangle).
\end{eqnarray}
Again, the state $|\psi_{kk}\rangle$ as well as all its components are symmetric with respect to fictitious indices enumerating bosons.

A remark concerning the notation used is in order here. The states $|1_k\rangle_A|1_l\rangle_B, |2_k\rangle_A|0\rangle_B,|0\rangle_A|2_k\rangle_B$ or $|1_k\rangle_A |1_k\rangle_B$ are separable with respect to the subsystems $A$ and $B$ in spite of their rather complicated form when written using original kets. This is because they are invariant under partial transposition with respect to the subsystem $A$ or $B$. Here partial transposition, say with respect to the subsystem $B$, simply amounts to exchanging indices between kets and bras having label $B$. For instance, $(|\phi^A_k\rangle|\phi^B_m\rangle\langle\phi^B_n|\langle\phi^A_l|)^{T_B} =
|\phi^A_k\rangle|\phi^B_n\rangle\langle\phi^B_m|\langle\phi^A_l|$. This ensures that after partial transposition the state remains symmetric with respect to exchange of indices enumerating bosons.

We clearly see the tensor product structure of two subsystems $A$ and $B$. After carrying out the trace over the subsystem $B$ we get
\begin{eqnarray}
&&\rho_A(kl) = p_{AA}(kl) |1_k,1_l\rangle_A {}_A\langle 1_k,1_l|+p_{BB}(kl)|0\rangle_A {}_A\langle 0|+p_A(k)p_B(l)|1_k\rangle_A{}_A\langle 1_k|+p_A(l)p_B(k)|1_l\rangle_A{}_A\langle 1_l|+\nonumber\\
&&\sqrt{p_A(k)p_B(l)p_A(l)p_B(k)}(|1_k\rangle_A{}_A\langle 1_l|+
|1_l\rangle_A{}_A\langle 1_k|)
\end{eqnarray}
for $k\neq l$ and
\begin{equation}
\rho_A(kk) =p_A^2(k)|2_k\rangle_A{}_A\langle 2_k|+p^2_B(k)|0\rangle_A{}_A\langle 0|+2p_A(k)p_B(k)|1_k\rangle_A{}_A\langle 1_k|
\end{equation}
for $k=l$.

Let us now discuss the consequences of the adopted definition for the amount of spatial entanglement. First of all, the states $|\psi_{kk}\rangle$ are entangled with the amount of entanglement given by the Shannon entropy of the binary probability distribution $p_A(k)^2,p_B(k)^2,2p_A(k)p_B(k)$. Please note that according to the incorrect reasoning presented before when one identifies subsystems with bosons themselves, there is no entanglement contained in this state. Similarly, the states with different $k$ and $l$ are not maximally entangled unless the effect of bunching disappears.

The amount of spatial entanglement between the regions $A$ and $B$ depends on the way the whole space available to the bosons is divided. For instance, if the bosons are trapped in a one dimensional harmonic trap a division of the whole space into two symmetric parts (with the demarcation line going through zero) yields a different amount of entanglement then an asymetric division. In the example of a one dimensional harmonic oscillator the symmetric split yields the largest amount of entanglement.

It is straightforward to extend the above discussion to many-boson systems.
We are not going to discuss this case here as in this paper we always deal with two bosons in the form of a reduced density matrix of the bigger many-boson system. Finally, we would like to point out that one can also define spatial entanglement in the sense specified for non-interacting fermions.

In the next part of the paper we study the properties of spatial entanglement between $A$ and $B$ as defined above in a system of non-interacting bosons in a grand canonical ensemble in a harmonic one dimensional trap.

\section{Spatial Entanglement in the Grand Canonical Ensemble}

We consider a system of non-interacting spinless bosons trapped in a one dimensional harmonic oscillator potential in thermal contact with a large system (reservoir). If we allow for the exchange of bosons between the trap and the reservoir the density matrix describing bosons inside the trap is called the grand canonical ensemble and reads
\begin{equation}
\rho = \frac{1}{Z_{\infty}} \sum_{n_0=0}^{\infty} \sum_{n_1=0}^{\infty} \dots \exp{ \left[-\frac{1}{\kappa T}
\left( \sum_{k=0}^{\infty} n_k E_k-\mu\sum_{k=0}^{\infty} n_k\right) \right] }
|n_0,n_1,\dots\rangle\langle n_0,n_1,\dots|,
\end{equation}
where $T$ is the temperature of the reservoir, $\kappa$ is Boltzmann constant, $E_k$ the energies of harmonic oscillator, $\mu$ chemical potential and the ket $|n_0,n_1,\dots\rangle$ describes a symmetrized state of $n_0$ bosons occupying the ground state, $n_1$ bosons occupying the first excited state etc. In all the calculations we put $\kappa = 1$ and $E_k = k$ ($k=0, 1,\dots$).

Due to the exchange of bosons between the trap and the reservoir the number of bosons in the trap fluctuates. To establish a link between experiments where the number of bosons in the trap is constant (such a situation is described by the canonical ensemble) one has to impose a condition that the mean number of bosons in the trap $\langle N\rangle$ is constant. This can be done by adjusting chemical potential with temperature. Unfortunately, to find a required dependence of $\mu$ on temperature one has to resort to numerical methods.

One can now apply the method presented in the previous section to discuss the spatial entanglement in the system. This is a very difficult task because of the large number of bosons in the trap (this number fluctuates around the mean value $\langle N\rangle$). However, one can ask an interesting question of how much of spatial entanglement is contained in an average pair of bosons in the trap. The amount of this spatial entanglement may be related to the phenomenon of the so-called off-diagonal long-range correlations, which indicate the onset of Bose-Einstein condensation (BEC) (or, in general, phase transitions) \cite{Yang}. In the system considered here (finite and one dimensional system of non-interacting bosons) one has to be careful how to define BEC because, strictly speaking, it does not exist. However, as discussed in Ref. \cite{ketterle} one can observe a phenomenon of the macroscopic occupation of the ground state, which may be taken as the definition of BEC.

Therefore, we focus our attention on the reduced density matrix $\rho(2)$ of an average pair of bosons in the trap
\begin{equation}
\rho(2) = \sum_{k} \frac{2 \langle n_k\rangle^2}{M} |\psi_{kk}\rangle\langle\psi_{kk}|+\sum_{k>l}\frac{\langle n_k\rangle\langle n_l\rangle }{M}|\psi_{kl}\rangle\langle\psi_{kl}|,
\label{two}
\end{equation}
where $M=\frac{1}{2}\left(3 \sum_k \langle n_k\rangle^2+\langle N\rangle^2\right)$, $\langle N\rangle=\sum_k\langle n_k\rangle$ is the mean number of particles in the trap and $\langle n_k\rangle = (\    exp{[\frac{1}{T}(E_k-\mu)]}-1)^{-1}$ is the mean number of bosons occupying the energy eigenket $|\phi_k\rangle$ of harmonic oscillator potential.

To find the amount of the spatial entanglement contained in $\rho(2)$ we could compute the so-called negativity introduced in Ref. \cite{werner}. The negativity is the sum of moduli of the negative eigenvalues of the partially transposed with respect to $B$ (or $A$ since the system is symmetric) density operator $\rho(2)$. Unfortunately, it is very difficult to diagonalize $\rho(2)^{T_B}$ since it is an infinite matrix. One would have to cut off the summations in (\ref{two}) and diagonalize it numerically but a reasonable cut-off leads to a huge matrix which dimension increases too fast to be tractable numerically. However, we can find an analytical formula for the lower bound for the total spatial entanglement contained in $\rho(2)$. This lower bound is related to the spatial entanglement related to a "coherent bunching" of bosons in either $A$ or $B$, i.e., to the states of the type $|1_k,1_l\rangle_A|0\rangle_B+|0\rangle_A|1_k,1_l\rangle_B$. To see this let us carry out the partial transposition of the matrix $\rho(2)$ with respect to $B$

\begin{eqnarray}
&&\rho(2)^{T_B} = \frac{1}{2 M}\sum_k \langle n_k\rangle^2\left(|2_k\rangle_A|2_k\rangle_B{}_A\langle 0|{}_B\langle 0|+|0\rangle_A|0\rangle_B{}_A\langle 2_k|{}_B\langle 2_k|\right)+\nonumber\\
&&\frac{1}{M}\sum_{k>l}p_{AA}(k,l)\langle n_k\rangle \langle n_l\rangle[\left( |1_k,1_l\rangle_A |1_k,1_l\rangle_B{}_A\langle 0|{}_B\langle 0| +
|0\rangle_A |0\rangle_B{}_A\langle 1_k,1_l|{}_B\langle 1_k,1_l|\right)+\Delta(2),
\label{pt}
\end{eqnarray}
where $\Delta(2)$ is orthogonal to the first two operators and is not, in general, a positive operator. Therefore, the negatives eigenvalues of the first two terms give us the lower bound for the total entanglement in $\rho(2)$. Taking into account the possible negative eigenvalues of $\Delta(2)$ can only increase the amount of total spatial entanglement.

To find the negative eigenvalues of the first two operators in (\ref{pt}) let us define a ket
\begin{equation}
|\chi\rangle = \sum_k \frac{1}{2}\langle n_k\rangle^2 |2_k\rangle_A |2_k\rangle_B+\sum_{k>l} \langle n_k\rangle\langle n_l\rangle |1_k,1_l\rangle_A|1_k,1_l\rangle_B.
\end{equation}
With this notation we have
\begin{equation}
\rho(2)^{T_B} = \frac{1}{M} (|\chi\rangle\langle 0|+|0\rangle\langle\chi|)+\Delta(2),
\end{equation}
where the ket $|0\rangle=|0\rangle_A|0\rangle_B$. The operator $|\chi\rangle\langle 0|+|0\rangle\langle\chi|$ has only one negative eigenvalue $\lambda$ given by
\begin{eqnarray}
&&\lambda=\frac{1}{M}\sqrt{\langle\chi|\chi\rangle} = \frac{1}{M}\left(\frac{1}{4}\sum_{k,l}\langle n_k\rangle^2 \langle n_l\rangle^2{}_A\langle 1_k|1_l\rangle_A^4+\right.\nonumber\\
&&\left. \sum_{k>l}\sum_{k'>l'}\langle n_k\rangle \langle n_l\rangle \langle n_{l'}\rangle \langle n_{k'}\rangle
{}_A\langle 1_k|1_{k'}\rangle_A^2{}_A\langle 1_l|1_{l'}\rangle_A^2p_{AA}(kl)p_{AA}(k'l')+\right.\nonumber\\
&&\left. \sum_{k'>l'}\sum_k \langle n_k\rangle^2 \langle n_{k'}\rangle\langle n_{l'}\rangle {}_A\langle 1_k|1_{k'}\rangle_A^2{}_A\langle 1_k|1_{l'}\rangle_A^2p_{AA}(k'l')\right),
\end{eqnarray}
where we have used the symmetry of the division into $A$ and $B$.
Therefore, the lower bound for the negativity of the state $\rho(2)$ equals to $\lambda$. Its behaviour with temperature is depicted in FIG. 1.
\begin{figure}[h]
\centering
\includegraphics[width=10cm]{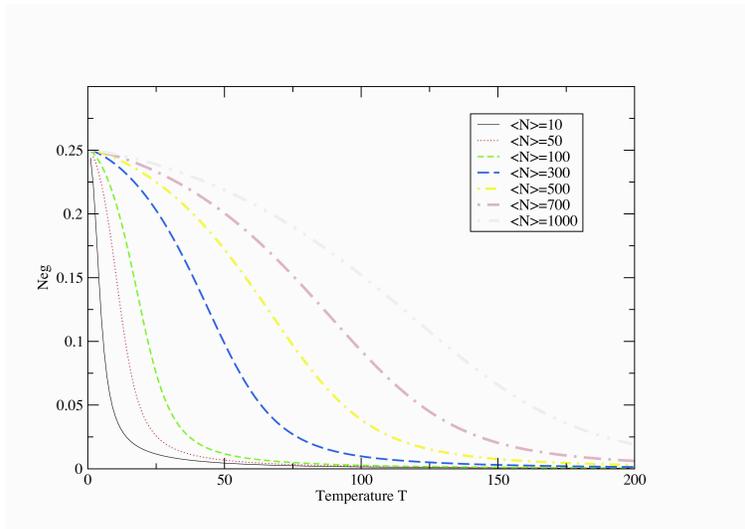}
\caption{Plot of the lower bound for the negativity of the state $\rho(2)$ (given by $\lambda$) versus temperature for the fixed mean value of the particles $\langle N\rangle$.}
\label{gc_mu}
\end{figure}

\section{Conclusions}

From the formula for $\lambda$ it can be easily shown that the spatial entanglement of an average pair of two bosons in the trap exists for arbitrary temperatures and it does not vanish with the increasing $\langle N\rangle$. This strongly suggests (see discussion in the Ref. \cite{vlatko-lunkes}) that the multipartite spatial entanglement of the bosons in the trap, i.e., the total spatial entanglement contained in the state $\rho$, should exist for arbitrary high temperatures.

It would be interesting to compute the total spatial entanglement
contained in the state $\rho(2)$ and investigate its behaviour
around the critical temperature for Bose-Einstein condensation (BEC)
understood as the macroscopic occupation of the ground state (see a
detailed discussion in the Ref. \cite{ketterle}). We conjecture that
there should be a relatively sharp decrease of the amount of spatial
entanglement contained in $\rho(2)$ around this temperature. This,
in turn, would suggest that there should be an analogous sharp
change of the total spatial entanglement contained in $\rho$.

As a final remark we would like to point out that entanglement
related to the "coherent bunching" of bosons in $A$ and $B$, i.e.,
the spatial entanglement contained in the states $|1_k,1_l\rangle_A
|0\rangle_B+|0\rangle_A|1_k,1_l\rangle_B$ is similar to the
so-called non-locality of a single photon \cite{singlephoton}.

\section{Acknowledgements}
We would like to thank V. Vedral and B.-G. Englert for useful
discussions. DK thanks Artur Ekert for illuminating discussions, MW
thanks \v{C}aslav Brukner and Artur Ekert. This work was supported
in part by the Singapore A*STAR Temasek Grant No. 012-104-0040 and
the MNiI Grant no. 1 P03 04927.

\end{document}